\documentclass{jpsj-suppl}
\usepackage{txfonts} 

\title{Towards an Experimental Determination of the Transition
Strength Between the Ground States of $^{20}$F and $^{20}$Ne}

\author{Oliver S.\ \textsc{Kirsebom}$^{1}$, 
Joakim \textsc{Cederk\"all}$^{2}$,
David G.\ \textsc{Jenkins}$^{3}$, 
Pankaj \textsc{Joshi}$^{3}$,
Rauno \textsc{Julin}$^{4}$, 
Anu \textsc{Kankainen}$^{4}$,
Tibor \textsc{Kib\'edi}$^{5}$,
Olof \textsc{Tengblad}$^{6}$,
Wladyslaw H.\ \textsc{Trzaska}$^{4}$}

\inst{$^{1}$Department of Physics and Astronomy, Aarhus University, DK-8000 Aarhus C, Denmark \\
$^{2}$Department of Physics, Lund University, SE-22100 Lund, Sweden \\
$^{3}$Department of Physics, University of York, York YO10 5DD, United Kingdom \\
$^{4}$University of Jyvaskyla, Department of Physics, P.O.\ Box 35, FI-40014 University of Jyvaskyla, Finland \\
$^{5}$Department of Nuclear Physics, The Australian National University, ACT 0200, Australia \\
$^{6}$Instituto de Estructura de la Materia, CSIC, Serrano 113 bis, E-28006 Madrid, Spain
}

\email{oliskir@phys.au.dk}

\recdate{June 23, 2011}

\abst{Electron capture on $^{20}$Ne is thought to play a crucial role in the final evolution of electron-degenerate ONe stellar cores. Recent calculations suggest that the capture process is dominated by the second-forbidden transition between the ground states of $^{20}$Ne and $^{20}$F, making an experimental determination of this transition strength highly desirable. %
To accomplish this task we are refurbishing an intermediate-image magnetic spectrometer capable of focusing 7~MeV electrons, and designing a scintillator detector surrounded by an active cosmic-ray veto shield, which will serve as an energy-dispersive device at the focal plane.}

\kword{$\beta$ decay, weak-interaction rates, nuclear experiment, stellar evolution}

\begin{document}
\maketitle



Stars in the mass range 8--10$M_{\odot}$ are believed to develop a core consisting mainly of O and Ne (the products of C burning) which will cool due to neutrino emission, causing the electrons to become strongly degenerate, while gradually acquiring more mass as nuclear burning continues in surrounding shells. %
If this situation persists for a sufficient period of time, the mass may reach the Chandrasekhar limit, at which point the ONe core will begin to collapse under its own gravitational weight. %
(In the case of a ONe white dwarf in a close binary system, stable mass transfer from the companion star could produce a similar result.) %
The ultimate fate of the core is, however, highly uncertain. %
As the density increases, so will the Fermi energy of the degenerate electrons, and eventually electron capture on $^{24}$Mg and, soon after, $^{20}$Ne will set in. The captures will act to reduce the electron pressure (thus accelerating the collapse), reduce the electron mole number and contribute to the heating of the core. Eventually, the temperature will become sufficiently high for $^{16}\text{O}+^{16}$O fusion to take place, causing a thermonuclear runaway (owing to the high degeneracy) which is believed to proceed as a sub-sonic burning front. The density at which oxygen ignites is crucial in determining what happens next: If the ignition density is below a certain critical value, the burning front will completely (or partially) disrupt the star. In the opposite case, the burning front will stall and the core will collapse into a neutron star~\cite{jones2016}. 
Since the first theoretical studies of ONe cores more than 30 years ago (see Ref.~\cite{miyaji1980} and references therein), the theoretical models have become increasingly sophisticated and improved atomic and nuclear data have become available. Yet the fate of stars in the mass range 8--10$M_{\odot}$ remains uncertain. 
Understanding the final evolution of these stars is an interesting problem in itself, but is also important from the perspective of nucleosynthesis: The birth and death rate of stars in the mass range 8--10$M_{\odot}$ is comparable to that of all stars heavier than 10$M_{\odot}$, and hence their contribution to galactic chemical evolution could be significant~\cite{wanajo09, wanajo13}. %
The problem has received renewed focus in recent years, see, {\it e.g.}, Refs.~\cite{poelarends08, takahashi13, jones13, jones2014, pinedo14, moeller2014, schwab2015, jones2016}. Of particular relevance to the present paper is the work of Mart\'inez-Pinedo {\it et al.}~\cite{pinedo14}, which has highlighted the potential importance of the second-forbidden transition between the ground states of $^{20}$Ne and $^{20}$F. This transition has previously been overlooked, but is likely to dominate the capture rate in an important temperature-density range. Subsequently, Schwab {\it et al.}~\cite{schwab2015} have investigated the impact of the second-forbidden transition on the ignition density, concluding that it is a major source of uncertainty in its determination, and hence, in determining if the collapse leads to a thermonuclear explosion or a neutron star (the other main source of uncertainty being the possible onset of semi-convection). 



\begin{figure}[tbh]
\includegraphics[width=0.49\linewidth, angle=0, clip=true, trim=90 100 80 90]{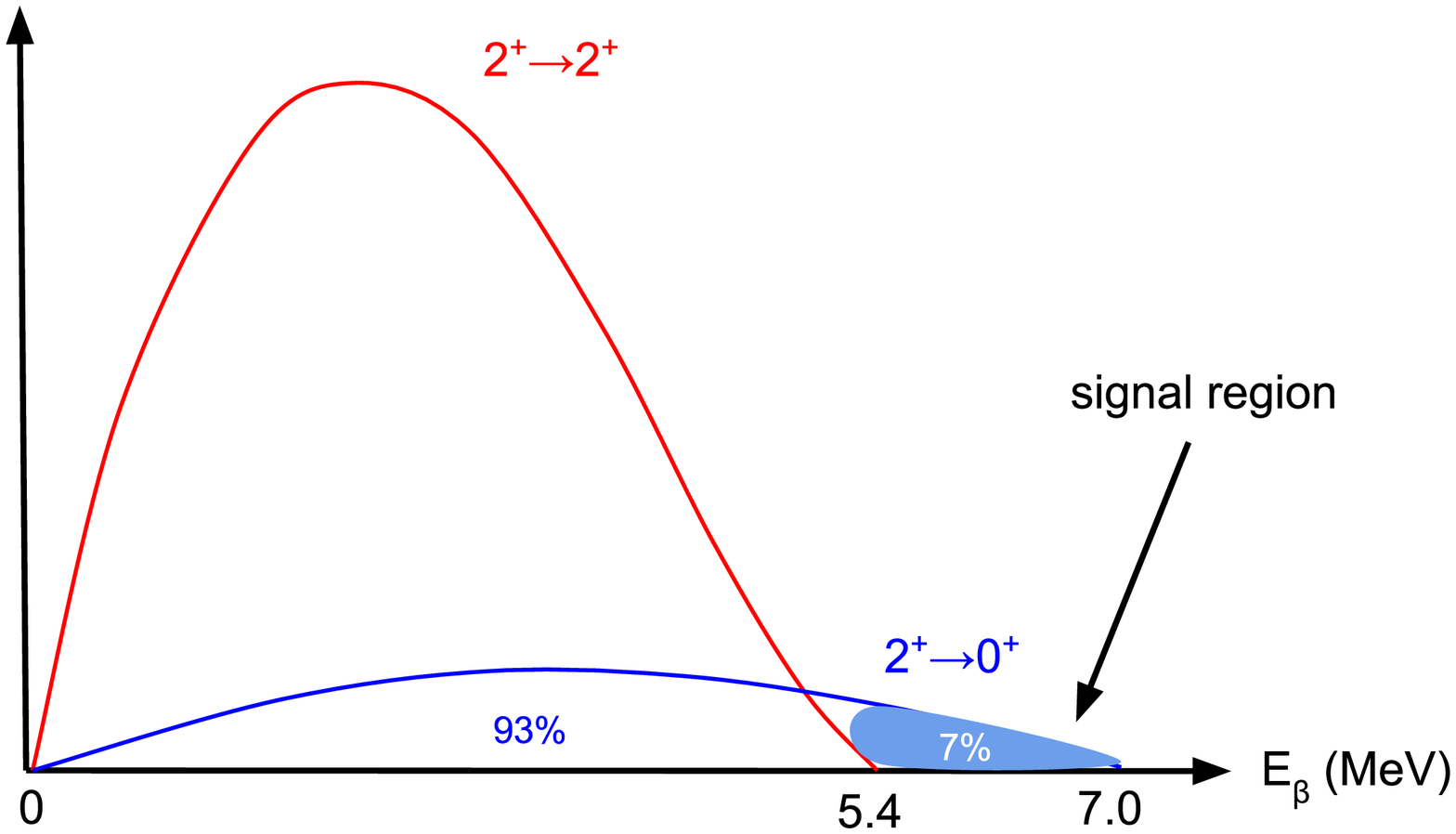}
\includegraphics[width=0.49\linewidth, angle=0, clip=true, trim=150 220 170 90]{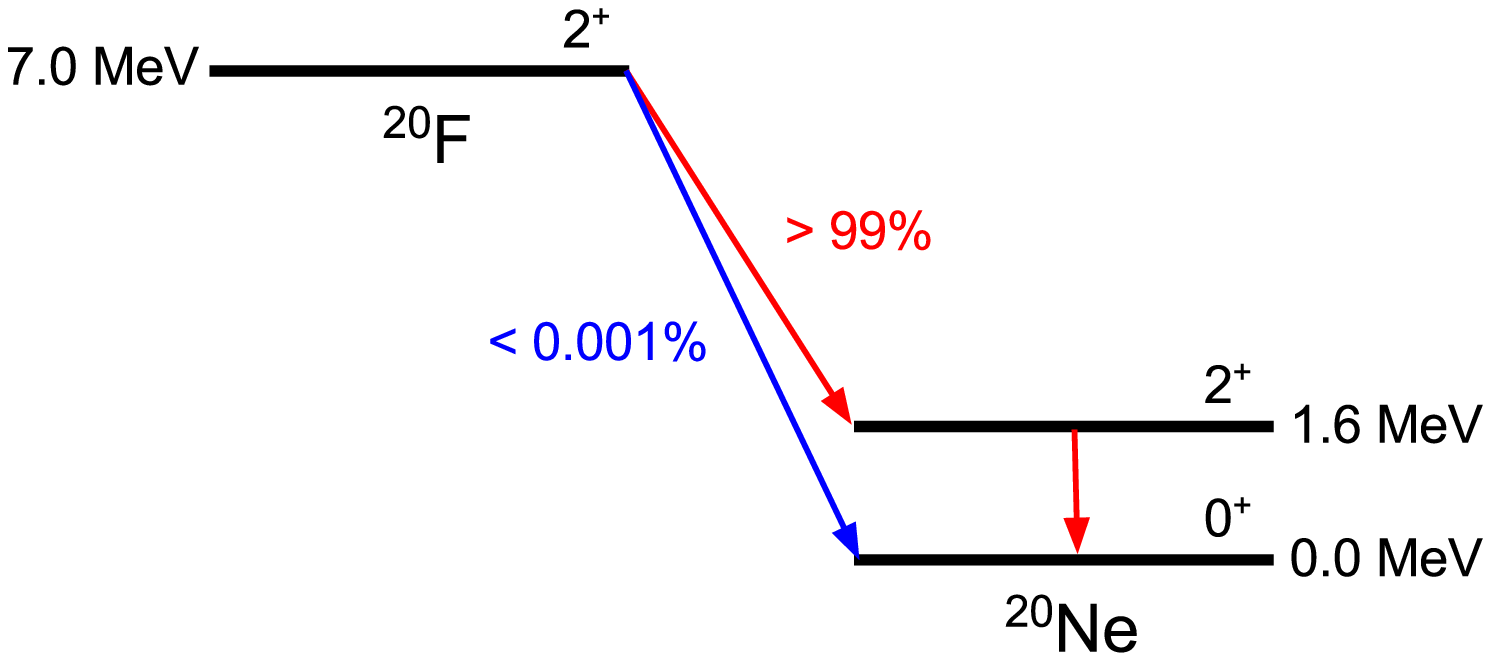}
\caption{Simplified decay scheme and schematic energy spectrum for the $\beta$ decay of $^{20}$F.}
\label{fig:decayscheme} 
\end{figure}

The strength of the second-forbidden transition is readily determined from the branching ratio (b.r.) of the inverse transition in the $\beta$ decay of $^{20}$F. The experimental determination of the b.r.\ is, however, not an easy task. As shown in Fig.~\ref{fig:decayscheme}, the decay proceeds mainly by an allowed $2^+\rightarrow 2^+$ transition to the first-excited state in $^{20}$Ne, situated 1.6~MeV above the ground state. This transition completely dominates the beta spectrum below $E_{\beta}=5.4$~MeV so the very weak, second-forbidden, $2^+\rightarrow 0^+$ transition can only be observed in the narrow energy range from 5.4~MeV to the end-point of the $\beta$ spectrum at 7.0~MeV. %
In 1954 Wong placed an upper limit of $\text{b.r.}<3\times 10^{-4}$ using a magnetic spectrometer~\cite{wong54}. In 1963 Glickstein and Winter obtained an upper limit of $\text{b.r.}<2\times 10^{-3}$ using a plastic-scintillator telescope~\cite{glickstein63}. Finally, in 1978 Calaprice and Alburger obtained an upper limit of $\text{b.r.}<1\times 10^{-5}$ using a magnetic spectrometer~\cite{calaprice78}, which is the lowest bound obtained to date. %
Note that the upper limits have been inferred under the assumption that the $\beta$ spectrum has the shape of an allowed transition. Deviations from the allowed shape can change the inferred branching ratio significantly and may change the astrophysical capture rate by a factor of 4--10~\cite{pinedo14}. It is therefore of interest to measure not only the integral of the $\beta$ spectrum above 5.4~MeV, but also the shape. %
(Assuming an allowed shape, the integral above 5.4~MeV is 7\% of the total integral. Using a more realistic shape, the result is 10\%~\cite{idini2014}.) %
A modern shell-model calculation predicts $\text{b.r.}=1.3\times 10^{-6}$~\cite{idini2014}, {\it i.e.}, one order of magnitude below the experimental upper limit, but it is not clear if this value can be trusted. The calculation is complicated due to the non-unique character of the $2^+\rightarrow 0^+$ transition. When applied to the $2^+\rightarrow 0^+$ transition between the ground states of $^{36}$Cl and $^{36}$Ar the model yields a strength that deviates substantially from the experimental value~\cite{idini_privcom}.


There are two aspects to the experimental determination of the b.r.\ of the $2^+\rightarrow 0^+$ transition: We must be able to produce $^{20}$F and we must be able to measure the high-energy end of the $\beta$ spectrum on-line (the half-life of $^{20}$F is 11~seconds). %
A straightforward and efficient way to produce $^{20}$F is through the $(d,p)$ reaction. This was the method used by Wong~\cite{wong54} and Calaprice and Alburger~\cite{calaprice78} while Glickstein and Winter~\cite{glickstein63} used the $(n,\gamma)$ reaction. The interactions between the beam and the target (typically, CaF$_2$) do, however, also produce other $\beta$-unstable isotopes besides $^{20}$F which act as a source of background. %
As an improvement upon the previous experiments, we will use the IGISOL-4 ion-guide system at the University of Jyv\"askyl\"a~\cite{moore2013} to transport the $^{20}$F ions to a separate experimental station, far removed from the prompt and delayed radiation at the target site. Furthermore, the ions will be bent trough a magnetic field whereby all unwanted $\beta$-unstable isotopes will be efficiently suppressed. With this approach the $^{20}$F yield at the experimental station is limited by the transmission efficiency of the ion guide. We have estimated that a yield of $0.4\times 10^{5}$ ions/s should be easily attainable. In comparison Calaprice and Alburger had $1\times 10^6$ ions/s. %

Even in a background-free environment the experiment remains challenging: Given the smallness of the b.r., the detection system must have high absolute efficiency for the experiment to be performed within a reasonable time. A plastic scintillator or a semiconductor detector situated close to the source can provide the required efficiency, but will also be sensitive to $\gamma$ rays, so $\beta\gamma$ summing at some rate is inevitable. In addition, there will be $\beta\beta$ pile-up. %
As a result the $\beta$ spectrum of the allowed transition will spill into the signal region between 5.4~MeV and 7.0~MeV, significantly limiting the experimental sensitivity. %
In principle, $\beta\gamma$ summing and $\beta\beta$ pile-up can be suppressed by employing an {\it array} of detectors, but our estimates suggest that in practice such an array would have to consist of a very large number of detectors ($>100$) situated at considerable distance from the source ($>1$~m). 
We have also considered alternative approaches to suppressing $\beta\gamma$ summing and $\beta\beta$ pile-up, including stacking the detectors to provide differential energy-loss information or using time-of-flight detectors or even Cherenkov detectors, but none of these approaches appear feasible. %

Instead we plan to use a setup consisting of a Siegbahn-Slatis type intermediate-image magnetic electron transporter combined with an energy-dispersive $\beta$ detector. (A similar approach was used by Calaprice and Alburger.) %
In this approach the mass-separated $^{20}$F beam is implanted in a thin foil. Two current-carrying coils arranged in a Helmholtz-type configuration generate a strong magnetic field ($B_{\text{max}}\approx 0.5$~T) that guides the most energetic electrons (those with energies between 5.4~MeV and 7.0~MeV) from the foil to the detector (situated on the symmetry axis at opposite ends of the setup), while a shield blocks the path of the $\gamma$ rays and the lesser energetic electrons. %
This setup was constructed at the University of Jyv\"askyl\"a in the 1980s and originally used for in-beam conversion-electron spectroscopy~\cite{julin1988}. Cooled silicon and germanium detectors were used as energy-dispersive devices because high resolution was desired. Since the present experiment does not require high resolution, we will use a plastic-scintillator detector, which has higher full-energy detection efficiency for electrons and lower $\gamma$ response, does not require cooling, thus simplifying the setup, and is more affordable. The expected resolution of the detector is 300~keV (full-width at half maximum) at 5~MeV~\cite{otto1979}. %
The detector will consist of a inner (signal) detector and an outer (veto) detector which will be used as an active shield against cosmic rays. %
The setup will also be equipped with a LaBr$_3$ detector which will provide absolute normalization by measuring the 1.6~MeV $\gamma$-ray intensity. %
Based on GEANT4 simulations the absolute detection efficiency of the setup is expected to be $\epsilon \approx 5$\%. Note that the magnetic field has a focusing effect, so the efficiency is significantly higher than the geometric solid angle of the detector ($\Omega/4\pi \approx 0.03$\%). %
Assuming a realistic yield of $0.4\times 10^{5}$ $^{20}$F ions/s and $\text{b.r.}=10^{-6}$, we expect $\sim 100$ counts inside the signal region in a 1-week long experiment, while the background due to the spill-over from main transition (due to the finite energy resolution, $\beta\gamma$ summing and $\beta\beta$ pile-up) will be $<10$ counts. %
In order to reduce the cosmic-ray induced background to a manageable level, {\it i.e.}, comparable to the signal of $\sim 100$ counts, the veto detector must achieve a rejection efficiency of $\sim 99$\% which appears realistic. %
If, contrary to expectations, the cosmic-ray induced background becomes a limiting factor, we may consider using a denser scintillator material such as LYSO which would allow for a thinner signal detector. (Such a detector would, however, have lower full-energy detection efficiency for electrons and higher $\gamma$ response.) %
The project is well underway. We expect to commission the setup in the fall of 2016 and perform the experiment during the following year.

\paragraph{Acknowledgments.}
We are grateful to Samuel Jones for making valuable suggestions upon reading the manuscript. This project is being supported by a grant from the Villum Foundation.


\begin{thebibliography}{10}

\bibitem{miyaji1980}
S.\ {Miyaji}, K.\ {Nomoto}, K.\ {Yokoi}, and D.\ {Sugimoto}: Pub.\
  Astron.\ Soc.\ Japan {\bf 32} (1980) 303.
\bibitem{jones2016}
S.\ Jones, {\it et al.}: Astron.\ Astrophys., in press;\, arXiv:1602.05771v2.
\bibitem{wanajo09}
S.\ Wanajo, K.\ Nomoto, H.-T.\ Janka, F.~S.\ Kitaura, and B.\ M{\"u}ller: Astrophys.\ J.\ {\bf 695} (2009) 208.
\bibitem{wanajo13}
S.\ Wanajo, H.-T.\ Janka, and B.\ M{\"u}ller: Astrophys.\ J.\ Lett.\
  {\bf 774} (2013) L6.
\bibitem{poelarends08}
A.~J.~T.\ Poelarends, F.\ Herwig, N.\ Langer, and A.\ Heger: Astrophys.\ J.\ {\bf 675} (2008) 614.
\bibitem{takahashi13}
K.\ Takahashi, T.\ Yoshida, and H.\ Umeda: Astrophys.\ J.\ {\bf 771} (2013) 28.
\bibitem{jones13}
S.\ Jones, {\it et al.}: Astrophys.\ J.\ {\bf 772} (2013) 150.
\bibitem{jones2014}
S.\ Jones, R.\ Hirschi, and K.\ Nomoto: Astrophys.\ J.\ {\bf 797} (2014) 83.
\bibitem{pinedo14}
G.\ Mart{\'i}nez-Pinedo, {\it et al.}: Phys.\ Rev.\ C {\bf 89} (2014) 045806.
\bibitem{moeller2014}
H.\ M{\"o}ller, S.\ Jones, T.\ Fischer, and G.\ Mart{\'i}nez-Pinedo: Proc.\ Sci.\ NIC XIII (2014) 125.
\bibitem{schwab2015}
J.\ Schwab, E.\ Quataert, and L.\ Bildsten: Mon.\ Not.\ R.\ Astron.\ Soc.\ {\bf 453} (2015) 1910--1927.
\bibitem{wong54}
C.\ Wong: Phys.\ Rev.\ {\bf 95} (1954) 761--764.
\bibitem{glickstein63}
S.~S.\ Glickstein and R.~G.\ Winter:  Phys.\ Rev.\ {\bf 129} (1963) 1281--1283.
\bibitem{calaprice78}
F.~P.\ Calaprice and D.~E.\ Alburger:  Phys.\ Rev.\ C {\bf 17} (1978) 730--738.
\bibitem{idini2014}
A.\ Idini, A.\ Brown, K.\ Langanke, and G.\ Martinez-Pinedo: Proc.\ Sci.\ NIC XIII (2014) 002.
\bibitem{idini_privcom}
A.\ Idini: private communication.
\bibitem{moore2013}
I. Moore, {\it et al.}: Nucl.\ Instr.\ Meth.\ B {\bf 317, Part B} (2013) 208--213.
\bibitem{julin1988}
R.\ Julin, {\it et al.}: Nucl.\ Instr.\ Meth.\ A {\bf 270} (1988) 74--77.
\bibitem{otto1979}
H.\ Otto, P.\ Peuser, G.\ Nyman, and E.\ Roeckl: Nucl.\ Instr.\ Meth.\ {\bf 166} (1979) 507--514.
\end{thebibliography}

\end{document}